\documentstyle[]{article}
\def\emline#1#2#3#4#5#6{%
       \put(#1,#2){\special{em:moveto}}%
       \put(#4,#5){\special{em:lineto}}}
\def\newpic#1{}
\author{A.V.Zabrodin\\
{\it Institute of Chemical Physics, Kosygina st. 4, 117334,}\\
{\it Moscow, Russia,}\\
{\it and}\\
{\it ITEP, 117259, Moscow, Russia,}\\
e-mail: zabrodin@vxitep.itep.ru}
\title{Quantum transfer matrices for discrete and continuous
quasi-exactly solvable problems}
\date{December 1994}
\begin{document}
\maketitle
\vspace{0.5cm}
\begin{abstract}
We clarify the algebraic structure of continuous and discrete
quasi-exactly solvable spectral problems by embedding them into
the framework of the quantum inverse scattering method. The quasi-exactly
solvable hamiltonians in one dimension are identified with traces
of quantum monodromy matrices for specific integrable systems with
non-periodic boundary conditions. Applications to the Azbel-Hofstadter
problem are outlined.
\end{abstract}

\section{Introduction} \medskip

At present time the methods related to quantum integrability are
highly developped. Originally they were invented and used in the context
of quantum field theory. However, the main ingredients proved to be purely
algebraic. They can be successfully applied to
quantum-mechanical (i.e., one-particle) problems as well.

An example of a new application of this kind is the recent progress
\cite{WZ1}, \cite{WZ2} in the famous problem of Bloch electrons in
magnetic field on a two-dimensional lattice \cite{Bemf} (sometimes called
the Azbel-Hofstadter problem). Even the one-particle problem is
non-trivial. In a proper gauge it reduces to a one-dimensional
quasiperiodic difference equation (Harper's equation is the most popular
example). It has been shown in \cite{WZ1}, \cite{WZ2} that some of these
equations admit partial exact solutions of the form typical for quantum
integrable systems: the eigenfunctions are
polynomials with the roots constrained by
Bethe equations. These solutions give some specific states (one for
each stable band), the energies being expressed through the roots in
a simple way. In the paper \cite{WZ3} this result has been generalized
to a whole class of second-order difference operators admitting partial
algebraization of the spectrum.

The form of the result suggests to ask for a direct connection with
quantum integrability. Indeed, such connection does exist.

A unified approach to quantum integrable systems is most elegantly
formulated in terms of the Quantum Inverse Scattering Method (QISM)
created by the Leningrad School \cite{F1} (for a more recent review
see \cite{F2}). In the paper \cite{WZ2} (Appendix B) it has been shown
how to embed the Azbel-Hofstadter problem into the QISM. The hamiltonian
has been identified with a quantum transfer matrix (trace of a quantum
monodromy matrix) for a specific integrable system with boundaries.
This gives a possibility to apply the powerful machinery of the QISM
such as functional Bethe ansatz \cite{SklLOMI}.

This paper may be considered as an extensive comment to Appendix B of
the paper \cite{WZ2}. We give a detailed construction of the quantum
monodromy matrices for the general family of difference equations
considered in \cite{WZ3}. In a more general context, we provide new formal
grounds for studying difference or differential operators in one variable
having a finite number of {\it polynomial} eigenfunctions. The continuum
limit corresponds to models with the rational $R$-matrix. In this case we
reproduce a class of second-order differential operators having the property
of the partial algebraization of the spectrum. Their eigenvalue equations
were considered in the literature some time ago \cite{ZU}, \cite{BV}
(the idea goes back to the papers \cite{Wint}, \cite{Gursey}). A systematic
treatment, based on the hidden dynamical $sl(2)$-symmetry,
was given in \cite{T1} (see also the reviews \cite{Sh}, \cite{T2} and
references therein). A different approach was suggested in \cite{Ush}.
These equations are known as "quasi-exactly solvable" problems ("quasi"
means that usually only a part of the spectrum can be found in a closed
algebraic form). The corresponding hamiltonians are known \cite{T1}
to be quadratic forms in the standard generators of $sl(2)$ (taken in
a finite-dimensional representation). In our approach the generators of
$sl(2)$ appear as matrix elements of the universal 2$\times $2 $L$-operator
of $XXX$-type. To $q$-deform this picture, one should use
$XXZ$-type $L$-operators. Their matrix elements are expressed through
generators of $U_q(sl(2))$, the $q$-deformation of the universal envelopping
algebra of $sl(2)$. This construction yields difference
equations~\footnote{An attempt to apply quantum algebras for generating
quasi-exactly solvable difference equations was made in \cite{OT}. However,
the authors used another version of the quantum algebra (which does not allow
one to construct hermitian hamiltonians; for details see \cite{WZ3}) and
did not discuss the Bethe ansatz solutions.}.

In short, we reduce the quasi-exactly solvable spectral problems mentioned
above to the problem typical for quantum integrable systems and lattice
statistical models, i.e., to diagonalization of a transfer matrix. Besides,
we give the QISM interpretation of isospectral
transformations of quasi-exactly
solvable hamiltonians (in the continuous case) under adjoint $SL(2)$-action.

Here is a more detailed description of the content.

In Sect.2 we describe universal trigonometric 2$\times $2 $L$-operators
depending on spectral parameter. There are two kinds of them: one is
related to $U_q(sl(2))$ (it is usually used in integrable $XXZ$ magnets
with higher spin), another one is associated to the dual algebra,
$A_q(SL(2))$. The matrix elements are expressed through the generators
of $U_q(sl(2))$ and $A_q(SL(2))$ respectively. The relevant representations
of these algebras are briefly reviewed. The rational limit of these
$L$-operators is also discussed. The former turns to the universal
$L$-operator of the isotropic $XXX$-type integrable magnet while the latter
becomes a $c$-number 2$\times $2 matrix independent of the spectral
parameter.

The necessary extraction from the formalism treating integrable systems
with boundaries is given in Sect.3. The starting point is "reflection
equations"~\cite{Chered}. Following \cite{SklJPhys}, we recall the
construction of quantum monodromy and transfer matrices for systems
with boundaries.

In Sect.4 we apply this general formalism to the elementary $L$-operators
and obtain in this way quasi-exactly solvable hamiltonians. Applications
to the Azbel-Hofstadter problem are outlined.

The rational (continuum) limit is treated in Sect.5. A comparison with the
representation in terms of Gaudin's magnet (suggested in \cite{Ush}) is
made. In Sect.6 we discuss isospectral transformations of continuous
quasi-exactly solvable hamiltonians under adjoint action of $SL(2)$.
The QISM interpretation of these transformations in terms of the rational
limit of the $L$-operator related to $A_q(SL(2))$ is suggested. Sect.7
contains conclusions and speculations on some open problems.
\medskip

\section{Elementary $L$-operators and quantum algebras}
\medskip

The standard basis of the quantum integrability is the Yang-Baxter
equation~\footnote{Sometimes (\ref{YBE1}) is called the Yang-Baxter
equation only if $T(u)=R(u)$, with considering it as a cubic functional
relation for $R(u)$.}
(YBE) with a spectral parameter $u$:
\begin{equation} \label{YBE1}
R(u/v)T_{1}(u)T_{2}(v)=T_{2}(v)T_{1}(u)R(u/v)\,,
\end{equation}
where $T_1 (u)=T(u)\otimes {\bf 1}$, $T_2 (u)={\bf 1}\otimes T(u)$.
We take $T(u)$ to be a 2$\times $2 matrix with operator matrix elements and
\begin{equation} \label{R}
R(u)= \left (
\begin{array}{cccc}
qu\! -\! q^{-1}u^{-1} & 0 & 0 & 0 \\
0 & u\!-\!u^{-1} & q\! -\! q^{-1} & 0 \\
0 & q\! -\! q^{-1} & u\! -\! u^{-1} & 0 \\
0 & 0 & 0 & qu\! -\! q^{-1}u^{-1} \end{array} \right )
\end{equation}
is the symmetric trigonometric $R$-matrix with the parameter $q$
(or its rational degeneration). In the trigonometric case we use the
multiplicative parametrization.

The equation (\ref{YBE1}) determines commutation relations for the
elements of the quantum monodromy matrix $T(u)$. The elementary solutions
of the YBE (i.e., those which can not be decomposed into a product of
simpler ones) are of particular importance for us. They are called
$L$-operators. In lattice integrable models (or spin chains) an
$L$-operator is usually associated to a lattice site. Here are two main
examples of $L$-operators.

{\it 1). $L$-operators associated to $U_q(sl(2))$.}

Consider the $L$-operator
\begin{equation} \label{Lax1}
L(u)=\left (
\begin{array}{cc}
uA-u^{-1}D & (q-q^{-1})C \\
(q-q^{-1})B & uD-u^{-1}A \end{array} \right )\,.
\end{equation}
It obeys the YBE (\ref{YBE1}) if and only if $A,\, B,\, C,\, D$
satisfy the commutation relations of the $U_q(gl(2))$ algebra~
\footnote{We use Koornwinder's notation \cite{K} for the generators.}
\cite{KR}, \cite{SklFunk}, \cite{D}, \cite{J}, \cite{RTF}~:
\begin{equation} \label{ABCD}
\begin{array}{l}
AB=qBA,\;\;\; BD=qDB\,, \\  \\
DC=qCD,\;\;\; CA=qAC\,, \\  \\
\!\! \phantom{a} [B,C]=\displaystyle{
\frac{A^2 -D^2 }{q-q^{-1} } }\,,\\  \\
\!\! \phantom{a} [A,D]=0\,.
\end{array}
\end{equation}
This quadratic algebra has two central elements. One of them is a $q$-analog
of the Casimir operator:
\begin{equation} \label{casimir1}
w=q^{-1}A^2 +qD^2 +(q-q^{-1})^2 BC\,,
\end{equation}
another one,
\begin{equation} \label{casimir2}
w_{0}=AD\,,
\end{equation}
for the $U_q(sl(2))$ case should be put equal to 1. If $q$ is a root of
unity some additional central elements appear.

Irreducible finite-dimensional representations of dimension $2j+1$ can
be expressed in the weight basis, where $A$ and $D$ are diagonal matrices:
$A= {\em diag}(q^{j},\ldots ,q^{-j})$. An integer or halfinteger $j$ is
spin of the representation. There exists the following realization
\cite{SklFunk} by difference operators acting in the linear space of
polynomials $F(z)$ of degree $2j$:
\begin{equation} \label{reprs}
\begin{array}{l}
AF(z)=q^{-j}F(qz)\,, \\  \\
BF(z)=-\displaystyle{ \frac{z}{q-q^{-1}} }
(q^{-2j}F(qz)-q^{2j}F(q^{-1}z))\,,\\  \\
CF(z)=\displaystyle{ \frac{1}{z(q-q^{-1})}} (F(qz)-F(q^{-1}z))\,, \\  \\
DF(z)=q^{j}F(q^{-1}z)\,.
\end{array} \end{equation}
Then $F_0(z)=1$ is the lowest weight vector whereas $F_{2j}(z)=z^{2j}$
is the highest weight vector, i.e., $CF_0 (z)=0$, $BF_{2j}(z)=0$.
The Casimir operator (\ref{casimir1}) in this realization is equal to
the $c$-number $q^{2j+1}+q^{-2j-1}$.

If $q$ is a root of unity there is, in addition, three-parametric
family of finite-dimensional representations having, in general, no
lowest and no highest weight \cite{SklFunk}. Sometimes they are called
{\it cyclic representations} \cite{Roche}. The difference quasi-exactly
solvable equations corresponding to the cyclic representations of
$U_q(sl(2))$ are particularly important \cite{WZ3} in applications to
the Azbel-Hofstadter problem. However, in this paper we do not consider
this case.

{\it 2). $L$-operators associated to $A_q(SL(2))$.}

Another important class of $L$-operators is constructed by means of the
dual quantum algebra $A_q(SL(2))$, the $q$-deformed algebra of functions
on the group $SL(2)$. Consider the operator matrix
\begin{equation} \label{Lax2}
\hat g(u) =\left (
\begin{array}{cc}
\hat a & u\hat b \\
u^{-1}\hat c & \hat d \end{array} \right )\,.
\end{equation}
It is easily verified that it satisfies the YBE (\ref{YBE1}),
\begin{equation} \label{YBE2}
R(u/v)\hat g_{1}(u)\hat g_{2}(v)=\hat g_{2}(v)\hat g_{1}(u)R(u/v)\,,
\end{equation}
if and only if $\hat a,\, \hat b,\, \hat c,\, \hat d$ obey the algebra
\begin{equation} \label{abcd}
\begin{array}{l}
\hat a \hat b =q\hat b \hat a\,,\;\;\; \hat b\hat d =q\hat d\hat b\,,\\  \\
\hat a\hat c =q\hat c\hat a\,,\;\;\; \hat c\hat d=q\hat d\hat c\,,\\  \\
\phantom{a}\!\! [\hat a,\hat d]=(q-q^{-1})\hat b\hat c\,,\\  \\
\phantom{a}\!\! [\hat b,\hat c]=0\,. \end{array}
\end{equation}
These are commutation relations for the generators of the dual algebra
of $U_q(gl(2))$ \cite{D}, \cite{Vaksman}.  We denote it $A_q(GL(2))$.
The conventional interpretation of $A_q(GL(2))$ identifies it with a
$q$-deformed algebra of functions on the group $GL(2)$.

The central element is $\hat b \hat c^{-1}$ (it belongs to an extended
algebra); another one is the $q$-determinant
$\hat a\hat d-q\hat b\hat c$, which for the $SL(2)$-case should be put
equal to 1. The corresponding factoralgebra is denoted $A_q(SL(2))$.
Restricting to the compact real form of the quantum group, one obtains
the algebra $A_q(SU(2))$, which was extensively studied \cite{Vaksman}.

For completeness, we give the list of irreducible unitary
representations of $A_q(SU(2))$ for real $q$ \cite{Vaksman}. There
are two series:

a). One-dimensional representations:
\begin{equation} \label{rep1}
\hat b=\hat c=0\,,\;\;\;
\hat a=\hat d^{-1}=e^{i\varphi }\,,
\end{equation}
$0\le \varphi <2\pi $.

b). Infinite-dimensional representations (parametrized by the same
"continuous spin" $\varphi $. They can be realized on functions in
one variable \cite{Soib}, \cite{Jurco}:
\begin{equation} \label{rep2}
\begin{array}{l}
\hat af(z)=-z^{-1}(f(qz)-f(q^{-1}z))\,, \\  \\
\hat bf(z)=q^{-1}e^{i\varphi }f(qz)\,, \\  \\
\hat cf(z)=-q^{2}e^{-i\varphi }f(qz)\,,\\  \\
\hat df(z)=qzf(qz)\,.
\end{array} \end{equation}
Note that $e^{i\varphi }$ enters (\ref{rep2}) in the same way as the
spectral parameter $u$ enters (\ref{Lax2}).

There are also some $c$-number solutions to (\ref{YBE1}): diagonal,
\begin{equation} \label{diag}
T(u)=\left ( \begin{array}{cc} * & 0 \\ 0 & * \end{array} \right )\,,
\end{equation}
and antidiagonal,
\begin{equation} \label{antidiag}
T(u)=\left ( \begin{array}{cc} 0 & * \\
\phantom{} * & 0 \end{array} \right )\,,
\end{equation}
where $*$ stands for an arbitrary $u$-independent $c$-number.
The former come from the $L$-operators $\hat g(u)$ taken in the
representation of the series a) (\ref{rep1}) while the latter correspond
to the $L$-operator (\ref{Lax1}) and the one-dimensional representation
($A=D=0$, $B$ and $C$ are $c$-numbers) of the algebra (\ref{ABCD}).

The trace (in the auxiliary two-dimensional space) of $T(u)$ obeying
(\ref{YBE1}) is a generating function of commuting integrals of motion:
$t(u)={\rm Tr}T(u)$,
\begin{equation} \label{commut1}
\phantom{a}[t(u),\, t(v)]=0\,.
\end{equation}
In the case of elementary $L$-operators there is only one independent
integral of motion (considered as a hamiltonian). Though the
commutativity (\ref{commut1}) is meaningless in this case, the transfer
matrix $t(u)$ possesses all necessary formal properties, which allow one
(at least, in principle) to apply the technique of the algebraic
(or functional) Bethe ansatz.

The rational limit means $q=e^{\hbar }$, $u=e^{\hbar \tilde u}$,
$\hbar \rightarrow 0$, and $\tilde u$ becomes an additive spectral
parameter. For future reference, let us present some formulas related
to the rational limit. In the rest of this section we write simply
$u$ instead of $\tilde u$.

The $L$-operator (\ref{Lax1}) becomes
\begin{equation} \label{Lax3}
L(u)=\left ( \begin{array}{cc} u+S_0 & S_{-} \\ S_{+} & u-S_0
\end{array} \right )\,, \end{equation}
where $S_i$ are generators of $sl(2)$:
\begin{equation} \label{sl2}
\phantom{a} [S_{\pm }, S_0 ]=\mp S_{\pm }\,,\;\;\;[S_{+},S_{-}]=2S_0\,.
\end{equation}
The correspondence with $U_q(sl(2))$ is as follows: $(A-D)/(2\hbar )
\rightarrow S_0$, $B\rightarrow S_{+}$, $C\rightarrow S_{-}$.

The realization (\ref{reprs}) is a smooth $q$-deformation of the
standard representation oof $sl(2)$ by first-order differential operators:
\begin{equation} \label{Dmod}
S_{-}=\frac{d}{dz}\,,\;\;S_0 =z\frac{d}{dz} -j\,,\;\;
S_{+}=-z^2 \frac{d}{dz} +2jz\,.
\end{equation}
This representation has been used to constract and classify linear
differential equations having polynomial solutions \cite{ZU}, \cite{BV},
\cite{T1}, \cite{Ush}. The QISM interpretation is given in Section 5.

The rational limit of (\ref{Lax2}) is simply a ($u$-independent) generic
"group element" of $SL(2)$  taken in the fundamental representation:
\begin{equation} \label{Lax4} g(u)= \left ( \begin{array}{cc}
 a & b \\ c & d \end{array}\right )\,. \end{equation}
Note that $a,\,b,\,c,\,d$ are $c$-numbers in this case since the algebra
(\ref{abcd}) becomes commutative. In Section 6 this "$L$-operator" is
used for a QISM interpretation of global $SL(2)$-rotations of
quasi-exactly solvable hamiltonians.
\medskip

\section{General properties of monodromy matrices for open integrable
spin chains} \medskip

Here we give a brief summary of the formalism treating integrable
systems with boundaries. The boundary conditions consistent with
integrability are determined by $c$-number 2$\times $2 matrices $K_l(u)$
and $K_r(u)$ (for the left and right boundary respectively) depending
on the spectral parameter and satisfying the "reflection equations"
(RE) \cite{Chered},
\begin{equation} \label{RE1}
\begin{array}{c}
R(u/v)(K_l (u)\otimes {\bf 1})R(uvq^{-1})({\bf 1}\otimes K_l (v)= \\  \\
=({\bf 1}\otimes K_l (v))R(uvq^{-1})(K_l (u)\otimes {\bf 1})R(u/v)\,,
\end{array} \end{equation}
\begin{equation} \label{RE2}
\begin{array}{c}
R(v/u)(K_{l}^{t}(u)\otimes {\bf 1})R(u^{-1}v^{-1}q^{-1})
({\bf 1}\otimes K_{r}^{t}(v))= \\  \\
({\bf 1}\otimes K_{r}^{t}(v))R(u^{-1}v^{-1}q^{-1})
(K_{r}^{t}(u)\otimes {\bf 1})R(v/u)
\end{array} \end{equation}
($t$ means the transposition) with the $R$-matrix (\ref{R}). Solutions
for $K_l$ and $K_r$ are related: if $K_l (u)$ is a solution to (\ref{RE1}),
then $K_{l}^{t}(u^{-1})$ is a solution to (\ref{RE2}). In the scattering
picture, the RE's describe the factorized scattering of a two-state
particle on left and right walls respectively.

More generally, one can consider operator solutions to the RE's
(\ref{RE1}), (\ref{RE2}) (i.e., matrices $K(u)$ with operator valued matrix
elements). Speaking informally, the wall may carry quantum numbers.
In this case the RE's determine commutation relations of the
matrix elements. They generate the "reflection algebra". In the case
of absence of the spectral parameter this algebra was studied in
\cite{Kulish}.

The QISM approach to integrable systems with boundaries was developped
by Sklyanin \cite{SklJPhys}. The main results of the paper \cite{SklJPhys}
are summarized below in the form of two theorems.

{\bf Theorem 1.} Let $T(u)$ satisfy the YBE (\ref{YBE1}) with the
$R$-matrix (\ref{R}) and let $K_l (u)$ (resp., $K_r (u)$) satisfy the RE
(\ref{RE1}) (resp., (\ref{RE2})) with the same $R$-matrix. Then
\begin{equation} \label{M1l}
{\cal K}_{l}(u)=T(u)K_l (u)\sigma _{2}T^t (u^{-1})\sigma _{2}\,,
\end{equation}
\begin{equation} \label{M1r}
{\cal K}_{r}(u)= \left ( T^{t}(u)K_{r}^{t}(u)\sigma _{2}T(u^{-1})
\sigma _{2}\right )^{t}.
\end{equation}
satisfy (\ref{RE1}) and (\ref{RE2}) respectively (here and below
$\sigma _{i}$ are Pauli matrices).

{\it Remark 1}. The theorem holds for both operator and $c$-number
$K$-matrices.

{\it Remark 2}. One may interpret (\ref{M1l}), (\ref{M1r}) as a
"dressing transformation": $K$ "dressed" by $T$ yields ${\cal K}$.

{\it Remark 3}. For unimodular $c$-number
matrices independent of $u$ the operation
$\sigma _{2}T^t \sigma _{2}$ is simply $T^{-1}$.

{\it Remark 4}. For a $c$-number matrix $K$, the (operator) matrix
${\cal K}$ is called the quantum monodromy matrix for an integrable
system with non-periodic boundary conditions.

It is convenient to represent (\ref{M1l}), (\ref{M1r}) pictorially
as follows:
$$
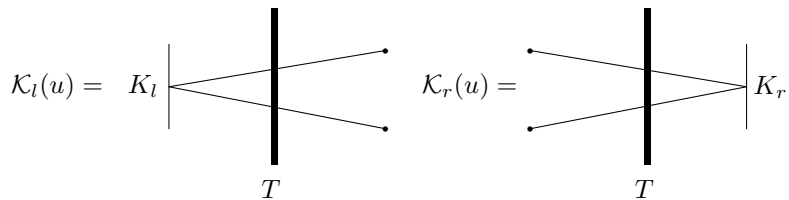
\begin{array}{l}
\unitlength=0.80mm
\special{em:linewidth 0.4pt}
\linethickness{0.4pt}
\begin{picture}(124.00,52.00)
\emline{24.00}{46.00}{1}{24.00}{32.00}{2}
\emline{24.00}{39.00}{3}{60.00}{45.00}{4}
\emline{24.00}{39.00}{5}{60.00}{32.00}{6}
\put(60.00,32.00){\circle*{1.00}}
\put(60.00,45.00){\circle*{1.00}}
\put(41.00,26.00){\rule{1.00\unitlength}{26.00\unitlength}}
\put(10.00,39.00){\makebox(0,0)[cc]{${\cal K}_{l}(u)=\ \ K_l$}}
\put(41.00,22.00){\makebox(0,0)[cc]{$T$}}
\emline{120.00}{32.00}{7}{120.00}{46.00}{8}
\put(84.00,45.00){\circle*{1.00}}
\put(84.00,32.00){\circle*{1.00}}
\put(103.00,26.00){\rule{1.00\unitlength}{26.00\unitlength}}
\put(103.00,22.00){\makebox(0,0)[cc]{$T$}}
\put(124.00,39.00){\makebox(0,0)[cc]{$K_r$}}
\put(74.00,39.00){\makebox(0,0)[cc]{${\cal K}_r(u)=$}}
\emline{120.00}{39.00}{9}{84.00}{45.00}{10}
\emline{84.00}{32.00}{11}{120.00}{39.00}{12}
\end{picture}
\end{array}
$$

{\bf Theorem 2.} Let ${\cal K}_{l}(u)$ and ${\cal K}_{r}(u)$ be any
solutions of (\ref{RE1}) and (\ref{RE2}) respectively. Then the
quantities
\begin{equation} \label{tau1}
\tau (u)={\rm Tr}( {\cal K}_{r}(u){\cal K}_{l}(u))
\end{equation}
form a commutative family:
\begin{equation} \label{commut2}
\phantom{a}[\tau (u),\,\tau (v)]=0\,.
\end{equation}

The quantity $\tau (u)$ is called a quantum transfer matrix. Its
diagonalization can be performed by means of the algebraic (or functional)
Bethe ansatz technique. To describe integrable open spin chains, one should
put ${\cal K}_{r}=K_r $ (a $c$-number solution) in (\ref{tau1})
and substitute ${\cal K}_{l}$ from (\ref{M1l}):

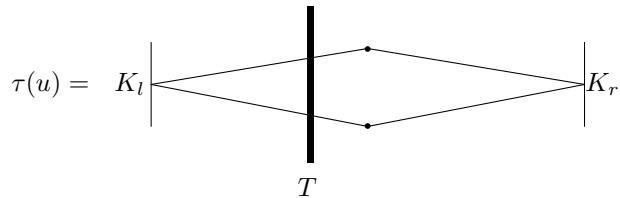
$$
\begin{array}{l}
\unitlength=0.80mm
\special{em:linewidth 0.4pt}
\linethickness{0.4pt}
\begin{picture}(94.00,52.00)
\emline{19.00}{46.00}{1}{19.00}{32.00}{2}
\emline{19.00}{39.00}{3}{55.00}{45.00}{4}
\emline{19.00}{39.00}{5}{55.00}{32.00}{6}
\put(55.00,32.00){\circle*{1.00}}
\put(55.00,45.00){\circle*{1.00}}
\put(45.00,26.00){\rule{1.00\unitlength}{26.00\unitlength}}
\put(18.00,39.00){\makebox(0,0)[rc]{$\tau (u)=\ \ K_l$}}
\put(45.00,22.00){\makebox(0,0)[cc]{$T$}}
\emline{91.00}{32.00}{7}{91.00}{46.00}{8}
\put(55.00,45.00){\circle*{1.00}}
\put(55.00,32.00){\circle*{1.00}}
\put(94.00,39.00){\makebox(0,0)[cc]{$K_r$}}
\emline{91.00}{39.00}{9}{55.00}{45.00}{10}
\emline{55.00}{32.00}{11}{91.00}{39.00}{12}
\end{picture}
\end{array}
$$
\medskip
\section{Trigonometric case}
\medskip

The boundary matrices for the reflection of a two-state particle
on a scalar wall are given by \cite{deVega}
\begin{equation} \label{Kleft}
K_l(u)=\left ( \begin{array}{ll}
2x_{0}(q^{-1}s^{-1}u-qsu^{-1}) \; & x_{+}(q^{-1}u^2 -qu^{-2}) \\
x_{-}(q^{-1}u^2 -qu^{-2}) \; & -2x_{0}(su-s^{-1}u^{-1})
\end{array} \right )\, ,
\end{equation} \medskip
\begin{equation} \label{Kright}
K_r (u)=\left( \begin{array}{ll}
2y_{0}(qtu-q^{-1}t^{-1}u^{-1}) \; & y_{+}(qu^2 -q^{-1}u^{-2}) \\
y_{-}(qu^2 -q^{-1}u^{-2}) \; & -2y_{0}(t^{-1}u-tu^{-1})
\end{array} \right )\, ,
\end{equation}
where $x_{0},\,x_{\pm},\,s$ and $y_{0},\,y_{\pm},\,t$ are arbitrary
parameters characterizing the boundary conditions.

The $L$-operator $L(u)$ (\ref{Lax1}) still satisfies the YBE (\ref{YBE1})
if $u$ is multiplied by a constant $k$. Substituting $L(uk)$ for $T(u)$
in (\ref{M1l}) we get the quantum monodromy matrix
\begin{equation} \label{M2}
M(u)=L(uk)K_{l}(u)\sigma _{2}L^{t}(u^{-1}k)\sigma _{2}
\end{equation}
(in this specific case we denote it $M(u)$).
The calculation of its matrix elements is straightforward. One should
take into account that the quadratic Casimir element $w$ (\ref{casimir1})
under any irreducible representation acts as a $c$-number. It is
convenient to represent the result in the following form.

Consider the operators
\begin{equation} \label{H1}
H_1 =x_{+}kAB+x_{-}k^{-1}CA+2(q-q^{-1})^{-1}x_{0}s^{-1}A^2 \,,
\end{equation}
\begin{equation} \label{H2}
H_2 =x_{+}k^{-1}DB+x_{-}kCD-2(q-q^{-1})^{-1}x_{0}sD^2\,,
\end{equation}
\begin{equation} \label{H3}
H_3\! =\!\!(q-q^{-1})^{-1}\!x_{+}(k^2 A^2 +k^{-2}D^2 )-(q-q^{-1})x_{-}C^2 +
2x_{0}(sk^{-1}\!DC\!-s^{-1}\!kAC),
\end{equation}
\begin{equation} \label{H4}
\bar H_{3}\!=\!\!(q-q^{-1})^{-1}\!x_{-}(k^{-2}A^2\! +k^2 D^2 )-
(q-q^{-1})x_{+}\!B^2\! +2x_{0}(skBD-s^{-1}\!k^{-1}\!BA\!).
\end{equation}
We note that $H_1 ,\,H_2 ,\,H_3 $ form a simple quadratic algebra
(a slightly different version of this algebra was previously studied
in \cite{GZh}):
\begin{equation} \label{algH1}
q^{-1}H_{\alpha}H_{\beta}-qH_{\beta}H_{\alpha}=g_{\gamma}H_{\gamma}+
h_{\gamma}\,,
\end{equation}
where $\{\alpha,\,\beta,\,\gamma \}$ stands for any cyclic permutation
of $\{1,\,2,\,3 \}$. The structure constants are:
\begin{equation} \label{gi}
g_1 =-(1+q^{-2})x_{+},\;\; g_2 =-(1+q^2 )x_{+},\;\; g_3 =(q+q^{-1})x_{-}\,,
\end{equation}
\begin{equation} \label{hi}
\begin{array}{l}
h_1 =2\displaystyle{ \frac{ x_{0}x_{+}}{q-q^{-1}}}
(s(k^2 +k^{-2})+q^{-1}s^{-1}w)\,,\\  \\
h_2 =-2\displaystyle{ \frac{ x_{0}x_{+} }{q-q^{-1}}}
(s^{-1}(k^2 +k^{-2})+qsw)\,, \\  \\
h_3 =\displaystyle{ \frac{1}{q-q^{-1}}}(4x_{0}^{2}-
x_{+}x_{-}(k^2 +k^{-2})w)\,.
\end{array} \end{equation}
The operators $H_1,\,H_2,\,\bar H_3 $ form a similar algebra; in particular,
\begin{equation} \label{algH2}
q^{-1}H_2 H_1 -qH_1 H_2 =(q+q^{-1})x_{+}\bar H_3 +h_3\,.
\end{equation}

Let us decompose $M(u)$ into the operator part $\hat M(u)$ and the
$c$-number part $M^{(c)}(u)$:
\begin{equation} \label{M3}
M(u)=\hat M(u) +M^{(c)}(u)\,.
\end{equation}
Then $\hat M(u)$ can be compactly written down in terms of the
operators (\ref{H1})-(\ref{H4}):
\begin{equation} \label{M4}
\hat M(u)=\frac{q^{-1}u^2 -qu^{-2}}{q-q^{-1}}
\left ( \begin{array}{cc}
-uH_1 +u^{-1}H_2 & H_3 \\
\bar H_3 & q^{-1}u^{-1}H_1 -quH_2
\end{array} \right )
\end{equation}
Note that the boundary parameters do not appear explicitly in (\ref{M4})
entering only through the structure constants of the algebra (\ref{algH1}).
For the $c$-number part one has
\begin{equation} \label{M5}
\begin{array}{l}
M^{(c)}_{11}(u)\!\!=\! M^{(c)}_{22}(qu^{-1}\!)\!=2x_{0}\displaystyle{
\frac{ w(su\!-\!s^{-1}u^{-1})\!+\!(k^2\! +\!k^{-2})
(q^{-1}s^{-1}u\!-\!qsu^{-1})}
{ (q-q^{-1})^{2} } }\,, \\  \\
M^{(c)}_{12}(u)=\displaystyle{\frac{x_{+}}{x_{-}}}M^{(c)}_{21}(u)=-x_{+}
\displaystyle{ \frac{ (u^2 +u^{-2})(q^{-1}u^2 -qu^{-2})}
{(q-q^{-1})^{2}} }\,.
\end{array} \end{equation}
It is clear from (\ref{H3}), (\ref{H4}) that the non-diagonal elements of
$M(u)$ generally do not have a zero mode (a "false vacuum") independent
of $u$. In such a case the general strategy of the functional Bethe
ansatz consists essentially in passing to the new basis formed by the
eigenvectors of $H_3$ or $\bar H_3$. Note that these operators contain
only elements of the lower (resp., upper) Borel subalgebra of $U_q(sl(2))$
(i.e, for example, $H_3$ is a quadratic form in $A$, $D$ and $C$, not $B$).
This allows one to find the eigenvectors of $H_3$ and $\bar H_3$ in a
quite explicit form \cite{WZ3}. Under the representation (\ref{reprs}) the
eigenfunctions are big $q$-Jacobi polynomials (see e.g. \cite{AW}).
This fact may be useful for diagonalization of $\tau (u)$ by means of
the functional Bethe ansatz.

Disregarding the $c$-number part, we get the quantum transfer matrix:
\begin{eqnarray} \label{tau2}
&\!\!\!\!&\!\!\! \tau (u)= {\em Tr}(K_r (u)\hat M(u))= \nonumber \\
&\!\!\!\!&\!\!\!=\!
\frac{ (qu^2\! -\!q^{-1}u^{-2})(q^{-1}u^2\! -\!qu^{-2})} {q-q^{-1} }
(2y_{0}(t^{-1}\! H_2\! -\! tH_1 )\! +\! y_{+}\bar H_3\! +\! y_{-}H_3\! ).
\end{eqnarray}
It is clear that in this case the family of commuting integrals of
motion generated by $\tau (u)$ contains only one (independent)
operator. In terms of $A,\, B,\, C,\, D$ (\ref{ABCD}) the transfer
matrix becomes a {\it generic homogeneous quadratic form} in these
operators (see (\ref{H1})-(\ref{H4})) since it depends on 7 parameters:
3 in each $K$-matrix (the common factor in (\ref{Kleft}) is inessential)
and $k$. Indeed, the total number of coefficients of a general
quadratic form is 10 but two of them contribute only to the $c$-number
term in (\ref{M3}) due to the two central elements ($AD=1$ and the
Casimir operator); besides, the common multiplier is also inessential.

As it is shown in \cite{WZ1}, \cite{WZ2}, the hamiltonian $\hat H$ of
the Bloch particle in a magnetic field is a particular quadratic form
in the $U_q(sl(2))$ generators with $|q| =1$ (the coefficients
depend on the gauge and the type of the lattice)
and therefore this system can be
considered as an integrable model. Here is a list of the most
important examples \cite{WZ1}, \cite{WZ2}, \cite{WZ3}.

1). {\it Square lattice, modified Landau gauge}: $K_l=\sigma _3$,
$K_r =\sigma _1$, $k\rightarrow \infty $,
\begin{equation} \label{ham1}
\hat H =-i(q-q^{-1})q^{-1/2}(CA+BD).
\end{equation}
The flux per plaquette is $\Phi =2\pi P/Q$ ($P$, $Q$ are coprime
integers), $q=e^{i\Phi /2}$.

2). {\it Square lattice, chiral gauge}: $K_l =\sigma _3 +
(q^{-1/2}u-q^{1/2}u^{-1})\sigma _{+}$, $K_l =\sigma _3 +
(q^{1/2}u-q^{-1/2}u^{-1})\sigma _{-}$, $k=1$,
\begin{equation} \label{ham2}
\hat H =i(q-q^{-1})q^{-1/2}(CA-BD+qBA-qCD).
\end{equation}
The flux per plaquette is $4\pi P/Q$ ($P$ odd).

3). {\it Triangular lattice, modified Landau gauge}: $x_{+}=y_{-}=0$,
$x_0 =y_0 =k=1$, $t=-s \rightarrow \infty $, $x_{-}=y_{+}
\rightarrow \infty $, $y_{+}/s\rightarrow 2\lambda q^{1/2}
\exp (-i\pi (P-1)/2)$, where $\lambda $ is the hopping amplitude
along the third axis, the flux per elementary triangle is
$\pi P/Q$ ($P$ odd). The hamiltonian is
\begin{equation} \label{ham3}
\hat H =\lambda (A^2 +D^2 )+e^{i\pi (P-1)/2}q^{-1/2}(q-q^{-1})
(CA+BD).
\end{equation}

4). {\it Square lattice, chiral gauge (a different version)}. We include
this example only for completeness. It is related to transfer matrix
for a {\it closed} chain: $t(u)={\em Tr}(\sigma _{2}L(u))$ (see
(\ref{commut1}). Here $L(u)$ is the $L$-operator (\ref{Lax1}) and
$\sigma _{2}$ is a $c$-number solution to (\ref{YBE1}) of the form
(\ref{antidiag}). The hamiltonian is
\begin{equation} \label{ham4}
\hat H=i(q-q^{-1})(C-B)\,,
\end{equation}
($Q$ odd, $P$ even).

Under the representation (\ref{reprs}) the transfer matrix (\ref{tau2})
becomes a second-order difference operator in $z$. Clearly, this
operator has the invariant subspace of polynomials spanned by
$1,\,z,\,z^2 ,\,\ldots ,z^{2j}$. The polynomial eigenfunctions lying
in this "algebraic" sector and the eigenvalues can be found in the
form standard for the algebraic Bethe ansatz technique. A detailed
analysis of the equations arising from (\ref{tau2}) and their Bethe
ansatz solutions is given in \cite{WZ3} (see also Appendix B
in \cite{WZ2}).

Here is the explicit form of these equations in terms of the
parameters of $K_l$ and $K_r$. By means of (\ref{reprs}) and
(\ref{H1})-(\ref{H4}) we rewrite the spectral problem for (\ref{tau2})
as follows:
\begin{equation} \label{spec1}
a(z)\psi (q^2 z)+d(z)\psi (q^{-2}z)-v(z)\psi (z)=E\psi (z)\,,
\end{equation}
where
\begin{equation} \label{a(z)}
a(z)\!\! =\!\! \left (\!\! q^{-2j+1}x_{+}z\!-\!\frac{2x_0}{sk}q^{-j}\! -\!
x_{-}k^{-2}\! z^{-1}\!\!
\right )\!\! \left (\!\! -q^{-2j}y_{+}z\!
+\! 2y_0 tkq^{-j}\! +\! q^{-1}y_{-}k^2
z^{-1} \phantom{\frac{q_0 }{s}}\!\!\!\!\!\!\!\!\right ), \end{equation}
\begin{equation} \label{d(z)} d(z)\!\!
=\!\! \left (\!\! q^{2j-1}x_{+}z\! -\! 2x_0
skq^{j}\phantom{\frac{q^j}{s}}\!\!\!\!\!\!-\! x_{-}k^2 z^{-1}\! \right )\!\!
 \left ( -q^{2j}y_{+}z\! +\! \frac{2y_0 }{tk}q^{j}\!
\! +qy_{-}k^{-2}z^{-1}\!  \right ),
\end{equation}
\begin{equation} \label{v(z)}
\begin{array}{l} v(z)=q^{2j}a(q^j z)+q^{-2j}d(q^{-j}z)- \\  \\
\displaystyle{-2(q^{j/2}-q^{-j/2})\left (
\left (y_0 x_{+}(\frac{q^{-j/2-1}}{tk}-q^{j/2+1}tk)
+y_{+}x_0 (q^{-j/2}sk-\frac{q^{j/2}}{sk})\right )z+\right.} \\  \\
\displaystyle{\left. +\left (y_0x_{-}(\frac{q^{j/2}k}{t}-
\frac{q^{-j/2}t}{k})+
y_{-}x_0 (\frac{q^{j/2+1}s}{k}-
\frac{q^{-j/2-1}k}{s})\right )z^{-1} \right )}.
\end{array} \end{equation}
Note that the dependence on the left and right boundary parameters
in $a(z)$ and $d(z)$ completely factorizes.

In general, some eigenfunctions of (\ref{tau2}) lie beyond the algebraic
sector. Hence this spectral problem is quasi-exactly solvable, i.e.,
only a part of the spectrum is available in a closed algebraic form.
There is one very important exception, where the spectrum itself is
finite and the algebraic sector totally covers it. This is
the case when $q$ is a root of
unity ($q=e^{i\pi P/Q}$, as before) and $j=(Q-1)/2$, so we obtain
an important class of periodic difference equations on a {\it finite}
ring-like lattice. Then all periodic eigenfunctions can be found
algebraically \cite{WZ3}. Furthermore, all quasiperiodic eigenfunctions
(corresponding to generic internal points of bands) are expressed in terms of
the family of the cyclic representations of $U_q(sl(2))$. This is just
the case relevant to the Azbel-Hofstadter problem. In particular,
for Example 3) above one obtains the spectral equation
\begin{equation} \label{spec2}
(z^{-1}-\lambda q)\psi (q^2 z)+(q^{-2}z-\lambda q^{-1})\psi (q^{-2}z)-
(z+z^{-1})\psi (z)=E\psi (z)\,.
\end{equation}
which provides the midband  points of the spectrum \cite{WZ3}. Another
approach to the Azbel-Hofstadter problem on a triangular lattice was
proposed in \cite{FK}.

Let us point out an alternative way to convert (\ref{tau2}) to a
difference operator. One may disregard the explicit formulas
(\ref{H1})-(\ref{H4}) connecting $H_i$ with $A,\,B,\,C,\,D$ and make
use of the representation theory of the algebra (\ref{algH1}),
(\ref{algH2}) developped in \cite{GZh}. Namely, the generators $H_1,\,
H_2,\,H_3,\,\bar H_3$ can be realized as second-order difference
operators of an essentially different form than the one following from
(\ref{H1})-(\ref{H4}) and (\ref{reprs}). It is evident from (\ref{tau2})
that under proper conditions the two forms are equivalent (at least in
the algebraic sector). More precisely, the corresponding operators, being
restricted to the algebraic sector, are connected by conjugation with
a certain matrix, i.e., by an isospectral transformation.  In some cases
such a transformation exists for the difference operator itself, not
only for its algebraic truncation. In this paper we will not discuss
this interesting question and only mention the following specific
example. The eigenfunctions of $H_3$ (\ref{H3}) under the representation
(\ref{reprs}) (where $A$ and $D$ are diagonal) are big $q$-Jacobi
polynomials (see e.g. \cite{AW}). On the other hand, as it follows from
\cite{GZh}, the eigenfunctions of $H_3$ in the basis, where $H_1$ is
diagonal are Askey-Wilson polynomials \cite{AW}. Our results indicate
that there should exist a similarity transformation between the two
difference operators. Some more details may be found in \cite{WZ3}.
\medskip

\section{Rational limit} \medskip

In this section we consider the limit $q\rightarrow 1$ (the "rational",
or continuum limit) of the monodromy matrix (\ref{M4}), providing a
basis for embedding the continuous quasi-exactly solvable problems
\cite{T1}, \cite{Ush} into the quantum inverse scattering approach.

The construction of the previous section may have different continuum
limits. Below we consider the most important one, which is directly
related to integrable models with the rational $R$-matrix. The rule
of performing this limit is as follows. Put $q=e^{\hbar }$,
$u=e^{\hbar \tilde u}$, $s=e^{\hbar \tilde s}$, $t=e^{\hbar \tilde t}$,
$k=e^{\hbar \tilde k}$, and find the $\hbar $-expansion of the monodromy
matrix (\ref{M2}) as $\hbar \rightarrow 0$, provided $x_0,\,x_{\pm },\,
y_0,\,y_{\pm}$ are $\hbar $-independent constants. In doing so, we will
write $u,\,s,\,t,\,k$ instead of $\tilde u,\,\tilde s,\,\tilde t,\,
\tilde k$ respectively. Since the trigonometric and rational cases never
mix, this convention can not lead to a confusion. Note that $u$ becomes
an additive spectral parameter.

This rule is equivalent to repeating the general arguments of Sects.~2 and
3 in the context of the rational
$R$-matrix and the $L$-operator (\ref{Lax3}).
The general $c$-number solutions to the rational RE are \cite{deVega}
\begin{equation} \label{Kleft1}
K_l =\left ( \begin{array}{cc} 2x_0 (u-s-1) & x_{+}(2u-1) \\
x_{-}(2u-1) & -2x_0 (u+s) \end{array} \right ),
\end{equation}
\begin{equation} \label{Kright1}
K_r =\left ( \begin{array}{cc} 2y_0 (u+t+1) & y_{+}(2u+1) \\
y_{-}(2u+1) & -2y_0 (u-t) \end{array} \right ).
\end{equation}
However, it is more convenient to proceed by considering the limits of
(\ref{M4}), (\ref{M5}).

The operators (\ref{H1})-(\ref{H4}) are expanded as
\begin{equation} \label{exp1}
\begin{array}{l}
H_1 = \hbar ^{-1}x_0 +H_{1}^{(0)}+\hbar H_{1}^{(1)}+{\cal O}(\hbar ^{2})\,,
\\  \\
H_2 =-\hbar ^{-1}x_0 +H_{2}^{(0)}+\hbar H_{2}^{(1)}+{\cal O}(\hbar ^{2})\,,
\\  \\
H_3 =\hbar ^{-1}x_{+}+2\hbar H_{3}^{(1)}+{\cal O}(\hbar ^{2})\,, \\  \\
\bar H_{3} =\hbar ^{-1}x_{-}+2\hbar \bar H_{3}^{(1)}+{\cal O}
(\hbar ^{2})\,,
\end{array} \end{equation}
where the operator coefficients are expressed through the generators
(\ref{sl2}) of $sl(2)$:
\begin{equation} \label{H10}
H_{1}^{(0)}=H_{2}^{(0)}=2x_0 S_0 +x_{+}S_{+}+x_{-}S_{-}-x_0 s\,,
\end{equation}
\begin{eqnarray} \label{H11}
&H_{1}^{(1)}=-H_{2}^{(1)}=&2x_0 S_{0}^{2}+x_{+}S_0 S_{+}+x_{-}S_{-}S_0-
\nonumber \\
&&-2x_0 sS_0 +x_{+}kS_{+}-x_{-}kS_{-}+\frac{1}{2}x_0 (s^2 -\frac{1}{3})\,,
\end{eqnarray}
\begin{equation} \label{H31}
H_{3}^{(1)}=x_{+}S_{0}^2 -x_{-}S_{-}^{2}-2x_0 S_0 S_{-}+2x_{+}kS_0+
2x_0 (s-k)S_{-}+x_{+}(k^2 -\frac{1}{12}),
\end{equation}
\begin{equation} \label{H41}
\bar H_{3}^{(1)}=x_{-}S_{0}^{2}-x_{+}S_{+}^{2}-2x_0 S_{+}S_0
-2x_{-}kS_0 +2x_0(s+k)S_{+}+x_{-}(k^2 -\frac{1}{12})\,.
\end{equation}
The operator $\hat M(u)$ (\ref{M4}) aquires a $c$-number part as
$\hbar \rightarrow 0$, with the leading term being singular ($\sim
\hbar ^{-1}$). In what follows we neglect all next-to-leading $c$-number
contributions since they are absolutely irrelevant. In particular, we
can throw away the $c$-number terms in (\ref{H10})-(\ref{H41}). Moreover,
it is easy to see that the $\hbar ^{-1}$-terms exactly cancel in the sum
(\ref{M3}). Finally, one obtains the following rational monodromy
matrix:
\begin{equation} \label{M6}
\begin{array}{c}
M^{(rational)}(u)=\displaystyle{ \lim _{\hbar \rightarrow 0}
\frac{1}{2\hbar }M(e^{\hbar u})}= \\  \\
=(2u-1)\displaystyle{ \left ( \begin{array}{cc}
-H_{1}^{(1)}\! -uH_{1}^{(0)}\,\, & H_{3}^{(1)} \\
\bar H_{3}^{(1)}\,\, & H_{1}^{(1)}\!\! -(u\!+\!1)H_{1}^{(0)}
\end{array} \right )}.
\end{array} \end{equation} Combining it with (\ref{Kright1}), we get the
transfer matrix:
\begin{eqnarray} \label{tau3}
\tau (u)&\!\!=\!\!&(4u^2 -1)\left ( y_{+}\bar H_{3}^{(1)}+y_{-}H_{3}^{(1)}-
2y_0 H_{1}^{(1)}-2y_0 t H_{1}^{(0)}\right )= \label{tau3prim} \\
&=&\!\!(4u^2 -1)\left ( (y_{+}x_{-}+y_{-}x_{+}-4y_0 x_0 )S_{0}^{2}-
y_{+}x_{+}S_{+}^{2}-y_{-}x_{-}S_{-}^{2}- \right. \nonumber \\
&&-(y_{+}x_0 +y_0 x_{+})(S_{+}S_0 +S_0 S_{+})-(y_{-}x_0 +y_0 x_{-})
(S_{-}S_0 +S_0 S_{-})+\nonumber \\
&&+2(k(y_{-}x_{+}-y_{+}x_{-})+2(s-t)y_0 x_0 )S_0 + \nonumber \\
&&+2((s+k+1/2)y_{+}x_0 -(t+k+1/2)y_0 x_{+})S_{+}+ \nonumber \\
&&\left. +2((s-k+1/2)y_{-}x_0
-(t-k+1/2)y_0 x_{-})S_{-}\phantom{S^2}\!\!\!\!\!\right ).
\end{eqnarray}
It is a generic mixed quadratic-linear form in $S_i$. The number of
independent parameters is the same as in (\ref{tau2}).

The diagonalization of (\ref{tau3}) gives (for the spin $j$ representation
(\ref{Dmod})) the following differential equation:
\begin{equation} \label{spec3}
\begin{array}{l}
\displaystyle{
-Q(z)\frac{d^2 \Psi (z)}{dz^2 }+\left ((j-\frac{1}{2})Q'(z)+2P(z)\right )
\frac{d\Psi (z)}
{dz}}\,- \\  \\
-\displaystyle{ \left ( \frac{1}{3}j(j-\frac{1}{2})Q''(z)+
2jP'(z)\right )\Psi (z)=E\Psi (z)}\,,
\end{array} \end{equation}
where
\begin{equation} \label{Q(z)}
Q(z)=(x_{+}z^2 -2x_0 z-x_{-})(y_{+}z^2 -2y_0 z-y_{-}),
\end{equation}
\begin{eqnarray} \label{P(z)}
P(z)&\!\!=&\!\!-\left ( (s+k+1/2)y_{+}x_0 -(t+k+1/2)y_0 x_{+}\right )z^2+
 \nonumber \\
&&+\left (k(y_{-}x_{+}-y_{+}x_{-})+2(s-t)y_0 x_0 \right )z+ \nonumber \\
&& +(s-k+1/2)y_{-}x_0 -(t-k+1/2)y_0 x_{-}\,.
\end{eqnarray}
Equations of this type are well-studied. If $Q(z)$ has 4 simple roots,
the eigenvalue problem (\ref{spec3}) can be reduced to Heun's equation
\cite{Erdel}. The transformation of (\ref{spec3}) to the Schr\"odinger
form is discussed in detail in the reviews \cite{Sh}, \cite{T2}.

A relation of (\ref{spec3}) to integrable spin chains was already
mentioned in the literature \cite{Ush}. However, the known relation
is absolutely different: the operator in the l.h.s. of (\ref{spec3})
is identified with the hamiltonian of inhomogeneous Gaudin's magnet
\cite{Gaudin} on 3 (or 4) sites with {\it periodic} boundary conditions.
It is known that Gaudin's magnet is a quasiclassical limit of the
integrable spin chain with the rational $R$-matrix.

In the present paper we identify (\ref{spec3}) with the eigenvalue
problem for the transfer matrix of a formal $XXX$-type "magnet" on
only one site~\footnote{This "spin chain" has only one spin because
(\ref{M2}) includes just one elementary $L$-operator.} but with
{\it non-periodic} boundary conditions. As it is shown in Sect.4, the
trigonometric generalization of this system leads to a class of
quasi-exactly solvable difference equations whereas an analog of the
Gaudin's magnet picture for the latter is not known.
\medskip

\section{Adjoint action of $SL(2)$} \medskip

There is an obvious group of isospectral transformations of (\ref{tau3}).
These transformations are induced by the adjoint action of $SL(2)$:
$S_i \rightarrow g^{-1}S_i g$. One may say that really different
spectral problems correspond to $SL(2)$-orbits in the space of quadratic
forms (\ref{tau3}). The generic orbit is 3-dimensional, so the number
of independent parameters is reduced to 4. Under the adjoint action of
a group element
\begin{equation} \label{g1}
g= \left (\begin{array}{cc} a & b \\ c & d \end{array} \right ), \;\;\;\;
ad-bc=1\,,
\end{equation}
the generators transform as follows:
\begin{equation} \label{adjoint}
\begin{array}{l}
S_{+}\rightarrow d^2 S_{+} +2cd S_0 -c^2 S_{-}\,, \\  \\
S_{0}\rightarrow bdS_{+}+(1+2bc)S_0 -acS_{-}\,, \\  \\
S_{-}\rightarrow -b^2 S_{+}-2abS_0 +a^2 S_{-}\,.
\end{array}
\end{equation}
Making this transformation in (\ref{tau3}), one gets an operator having the
same spectrum for any spin $j$. Our aim in this section is to show how
this transformation may be interpreted in terms of the QISM.

To do this, recall the $c$-number solutions (\ref{Lax4}) of the rational
YBE. Let us consider the group element (\ref{g1}) as such a solution and
apply Theorem 1 (see Sect.3) to the pair $g,\,K_l$ (or $g,\,K_r$) for
$K_l$ ($K_r$) given by (\ref{Kleft1}) ((\ref{Kright1})). It is convenient
to fix common multipliers in (\ref{Kleft1}) and (\ref{Kright1}) by putting
$(2s+1)x_0 =1$, $(2t+1)y_0 =1$. Then $K_l$ (resp., $K_r$) depends on a
3-dimensional vector ${\bf x}=(x_1 ,\, x_2 ,\, x_0 )$ (resp., ${\bf y}=
(y_1 ,\, y_2 ,\, y_0 )$), where $x_{\pm }=x_1 \pm x_2$, $y_{\pm }=
y_1 \pm y_2$~. Working in this normalization, we indicate the dependence
on ${\bf x}$ and ${\bf y}$ explicitly:
\begin{equation} \label{Kleft2}
K_l (u;{\bf x})=(2u-1)({\bf x}{\bf {\sigma}})-{\bf 1}\,,
\end{equation}
\begin{equation} \label{Kright2}
K_r (u;{\bf y})=(2u+1)({\bf y}{\bf {\sigma}})+{\bf 1}\,,
\end{equation}
where $({\bf x}{\bf {\sigma}})=x_1 \sigma _{1}+x_2 \sigma _{2}+
x_0 \sigma _{3}$ denotes the inner product of 3-dimensional vectors
(here $\sigma _{i}$ are Pauli matrices), and ${\bf 1}$ is the unit
matrix.

Now, according to Theorem 1, we should "dress" $K$-matrices using
(\ref{M1l}), (\ref{M1r}). In the simple case at hand the dressing is
reduced to the conjugation:
\begin{equation} \label{Kleft3}
\begin{array}{l}
\unitlength=1.00mm
\special{em:linewidth 0.4pt}
\linethickness{0.4pt}
\begin{picture}(90.00,53.00)
\emline{8.00}{46.00}{1}{8.00}{32.00}{2}
\emline{8.00}{39.00}{3}{44.00}{45.00}{4}
\emline{8.00}{39.00}{5}{44.00}{32.00}{6}
\put(44.00,32.00){\circle*{1.00}}
\put(44.00,45.00){\circle*{1.00}}
\put(4.00,39.00){\makebox(0,0)[cc]{$K_l$}}
\put(25.00,22.00){\makebox(0,0)[cc]{$g$}}
\emline{25.00}{45.00}{7}{25.00}{42.00}{8}
\emline{25.00}{41.00}{9}{25.00}{38.00}{10}
\emline{25.00}{37.00}{11}{25.00}{34.00}{12}
\emline{25.00}{33.00}{13}{25.00}{30.00}{14}
\emline{25.00}{29.00}{15}{25.00}{26.00}{16}
\emline{25.00}{46.00}{17}{25.00}{49.00}{18}
\emline{25.00}{50.00}{19}{25.00}{53.00}{20}
\put(49.00,39.00){\makebox(0,0)[lc]{$=gK_l(u;{\bf x})g^{-1}=K_l(u;g{\bf x})$,}}
\end{picture}
\end{array}
\end{equation}
\begin{equation} \label{Kright3}
\begin{array}{l}
\unitlength=1.00mm
\special{em:linewidth 0.4pt}
\linethickness{0.4pt}
\begin{picture}(90.00,53.00)
\put(3.00,32.00){\circle*{1.00}}
\put(3.00,45.00){\circle*{1.00}}
\emline{39.00}{32.00}{1}{39.00}{46.00}{2}
\put(3.00,45.00){\circle*{1.00}}
\put(3.00,32.00){\circle*{1.00}}
\put(42.00,39.00){\makebox(0,0)[lc]{$K_r=g^{-1}K_r(u;{\bf y})g=K_r(u;g^{-1}{\bf
y})$,}}
\emline{39.00}{39.00}{3}{3.00}{45.00}{4}
\emline{3.00}{32.00}{5}{39.00}{39.00}{6}
\emline{19.00}{53.00}{7}{19.00}{50.00}{8}
\emline{19.00}{49.00}{9}{19.00}{46.00}{10}
\emline{19.00}{45.00}{11}{19.00}{42.00}{12}
\emline{19.00}{41.00}{13}{19.00}{38.00}{14}
\emline{19.00}{37.00}{15}{19.00}{34.00}{16}
\emline{19.00}{33.00}{17}{19.00}{30.00}{18}
\emline{19.00}{29.00}{19}{19.00}{26.00}{20}
\put(19.00,23.00){\makebox(0,0)[cc]{$g$}}
\end{picture}
\end{array}
\end{equation}
where the dashed lines (carrying trivial one-dimensional "quantum space")
denote the insertions of the "$L$-operator" $g$.
The equalities immediately follow from (\ref{Kleft2}), (\ref{Kright2}).
The shorthand $g{\bf x}$ means the adgoint action of $g$ to the 3-component
vector ${\bf x}$. One concludes from (\ref{Kleft3}), (\ref{Kright3})
that the dressing in the case at hand is equivalent to a rotation of the
vector parameter.

Let us represent the generators $S_i$ as a vector with
operator-valued components:
\begin{equation} \label{Svec}
{\bf S}=(\frac{1}{2}(S_{-}+S_{+}),\,\frac{1}{2i}(S_{-}-S_{+}),\,
S_0 ).
\end{equation}
The adjoint action $g^{-1}{\bf S}$ is given by (\ref{adjoint}). In these
terms the transfer matrix (\ref{tau3}) can be written in the form:
\begin{eqnarray} \label{tau4}
\tau (u)&\!\!=&\!\!(4u^2 -1)\left (\phantom{S^2}\!\!\!\!\!\!
(({\bf y}\times {\bf S})({\bf x}\times {\bf S}))+
(({\bf x}\times {\bf S})({\bf y}\times {\bf S}))-
({\bf y}{\bf S})({\bf x}{\bf S})- \right. \nonumber \\
&\!\!&\left. \phantom{S^2}\!\!\!\!\!
-({\bf x}{\bf S})({\bf y}{\bf S})+2(({\bf y}-{\bf x}){\bf S})+
4ik(({\bf y}\times {\bf x}){\bf S})\right ) +\:\:
\mbox{$c$-number}\,,
\end{eqnarray}
where ${\bf a}\times {\bf b}$ denotes the skew product of 3-vectors:
$({\bf a}\times {\bf b})_{\alpha }=\epsilon _{\alpha \beta \gamma}
a_{\beta }b_{\gamma }$~. It is clear from (\ref{tau4}) that the operator
part of $\tau (u)$ is invariant under simultaneous rotations of all the
vectors ${\bf x}$, ${\bf y}$ and ${\bf S}$. In other words, the rotation
$g^{-1}{\bf S}$ (\ref{adjoint}) in (\ref{tau4}) is equivalent to
${\bf x}\rightarrow g{\bf x}$, ${\bf y}\rightarrow g{\bf y}$ given by the
"dressing" (\ref{Kleft3}), (\ref{Kright3}). The dressing means the
insertion of $g$ to the left and $g^{-1}$ to the right of the line
corresponding to $L(u)$ (\ref{Lax3}). Schematically,
\begin{equation}\label{pict3}
\begin{array}{l}
\unitlength=0.50mm
\special{em:linewidth 0.4pt}
\linethickness{0.4pt}
\begin{picture}(196.00,53.00)
\emline{7.00}{46.00}{1}{7.00}{32.00}{2}
\emline{7.00}{39.00}{3}{43.00}{45.00}{4}
\emline{7.00}{39.00}{5}{43.00}{32.00}{6}
\put(43.00,32.00){\circle*{1.00}}
\put(43.00,45.00){\circle*{1.00}}
\put(5.00,39.00){\makebox(0,0)[rc]{$K_l$}}
\put(39.00,22.00){\makebox(0,0)[cc]{$L^{(g)}$}}
\emline{79.00}{32.00}{7}{79.00}{46.00}{8}
\put(43.00,45.00){\circle*{1.00}}
\put(43.00,32.00){\circle*{1.00}}
\put(101.00,39.00){\makebox(0,0)[cc]{$K_r\ \ =\ \ K_l$}}
\emline{79.00}{39.00}{9}{43.00}{45.00}{10}
\emline{43.00}{32.00}{11}{79.00}{39.00}{12}
\put(39.00,26.00){\rule{1.00\unitlength}{26.00\unitlength}}
\emline{121.00}{46.00}{13}{121.00}{32.00}{14}
\emline{121.00}{39.00}{15}{157.00}{45.00}{16}
\emline{121.00}{39.00}{17}{157.00}{32.00}{18}
\put(157.00,32.00){\circle*{1.00}}
\put(157.00,45.00){\circle*{1.00}}
\put(153.00,22.00){\makebox(0,0)[cc]{$L$}}
\emline{193.00}{32.00}{19}{193.00}{46.00}{20}
\put(157.00,45.00){\circle*{1.00}}
\put(157.00,32.00){\circle*{1.00}}
\put(196.00,39.00){\makebox(0,0)[lc]{$K_r$}}
\emline{193.00}{39.00}{21}{157.00}{45.00}{22}
\emline{157.00}{32.00}{23}{193.00}{39.00}{24}
\put(153.00,26.00){\rule{1.00\unitlength}{26.00\unitlength}}
\emline{139.00}{53.00}{25}{139.00}{50.22}{26}
\emline{139.00}{49.00}{27}{139.00}{46.22}{28}
\emline{139.00}{45.00}{29}{139.00}{42.22}{30}
\emline{139.00}{41.00}{31}{139.00}{38.22}{32}
\emline{139.00}{37.00}{33}{139.00}{34.22}{34}
\emline{139.00}{33.00}{35}{139.00}{30.22}{36}
\emline{139.00}{29.00}{37}{139.00}{26.22}{38}
\emline{170.00}{53.00}{39}{170.00}{50.22}{40}
\emline{170.00}{49.00}{41}{170.00}{46.22}{42}
\emline{170.00}{45.00}{43}{170.00}{42.22}{44}
\emline{170.00}{41.00}{45}{170.00}{38.22}{46}
\emline{170.00}{37.00}{47}{170.00}{34.22}{48}
\emline{170.00}{33.00}{49}{170.00}{30.22}{50}
\emline{170.00}{29.00}{51}{170.00}{26.22}{52}
\put(170.00,22.00){\makebox(0,0)[cc]{$g^{-1}$}}
\put(139.00,22.00){\makebox(0,0)[cc]{$g$}}
\end{picture}
\end{array}
\end{equation}

Another way to see this is to observe that
\begin{equation} \label{transform}
g^{-1}L(u+k)g=(u+k){\bf 1}+({\bf {\sigma}}(g^{-1}{\bf S}))
\end{equation}
for the $L$-operator (\ref{Lax3}), and $\sigma _{2}L^{t}(-u)\sigma _{2}=
-L(u)$~. Then
\begin{equation} \label{tau5}
\tau (u)=-{\rm Tr}\left ( K_r (u;{\bf y})L(u+k)K_l (u;{\bf x})
L(u-k)\right ),
\end{equation}
and the transformation ${\bf S}\rightarrow g^{-1}{\bf S}$ in $L(u\pm k)$
leads to
\begin{eqnarray} \label{tau6}
\tau ^{(g)}(u)&\!\!=&\!\!-{\rm Tr}\left (K_r (u;{\bf y})g^{-1}
L(u+k)g K_l (u;{\bf x})g^{-1}L(u-k)g\right )= \nonumber \\
&\!=&\!\!-{\rm Tr}\left ( K_r (u;g{\bf y})L(u+k)K_l (u;g{\bf x})L(u-k)\right
), \end{eqnarray} which is equivalent to (\ref{tau4}) due to (\ref{Kleft3}),
(\ref{Kright3}).

It is an interesting open problem to find a proper $q$-analog of the
transformation considered in this section. In particular, it is not known
whether there are any isospectral subfamilies among the operators of
the form (\ref{tau2}) other than the trivial ones ($B \rightarrow
e^{-2i\varphi}B$, $C\rightarrow e^{2i\varphi }C$), which correspond to the
similar insertion of $\hat g(u)$ (\ref{Lax2}) taken in the one-dimensional
representation (\ref{rep1}). We hope that our approach may help to
solve this problem.
\medskip

\section{Concluding remarks} \medskip

We have shown that both discrete and continuous quasi-exactly solvable
problems of quantum mechanics are tractable in the framework of the
quantum inverse scattering method. Quantum transfer matrices for a
peculiar simple integrable system with boundaries yield the complete
collection of quasi-exactly solvable hamiltonians. These hamiltonians
are quadratic forms in the generators of $U_q(sl(2))$ (or $sl(2)$ in the
rational limit) taken in a finite-dimensional representation.

This reformulation opens a way to apply the powerful methods specific
for quantum integrable systems. For the representations of $U_q(sl(2))$
having both highest and lowest weights these methods give
the results which are eigther already known or can be obtained by means of
more elementary tools \cite{WZ3}. However, if $q$ is a root of unity
there exists a family of cyclic representations having in general
neigther highest nor lowest weights. This is just the case relevant to
the Azbel-Hofstadter problem, where the generic points of bands are
described by cyclic representations. The Bethe ansatz in this case is
much harder problem. Here the reformulation in terms of the QISM may led
to really new outcomes. The appropriate method is the technique of
Baxter's intertwining vectors applied to the chiral Potts model
(which is also known to be connected with cyclic representations)
in \cite{BS}. Recently, this method was applied \cite{FK} to the
Azbel-Hofstadter problem. Presumably, the method should work for any
operator of the form (\ref{tau2}) as well.

Let us point out two questions, where the results of the present paper
may contribute something to the conceptual understanding.

One of them was already mentioned at the end of Sect.6. It is the question
about isospectral subfamilies among operators of the form (\ref{tau2})
(or, rather, about a proper $q$-analog of them). In the continuous case,
they are orbits of the adjoint action of $SL(2)$ group elements in the
space of quadratic forms in $sl(2)$-generators. We have seen that the
group elements may be considered as $L$-operators (obeying the YBE with
spectral parameter), the adjoint action being an insertion of them into
the monodromy matrix. Remarkably, each ingredient of this picture has
its natural $q$-deformed counterpart. The notion of a quantum group-like
element was recently discussed \cite{qtau} (from another point of view)
in connection with quantum $\tau $-functions. A comparison of these
studies with our results may be fruitful for both approaches.

A related question concerns separation of variables. It is known \cite{Kuz}
that inhomogeneous $n$-site Gaudin's magnets are in one-to-one
correspondence with separated coordinate systems for the Laplace-Beltrami
operator on the $(n-1)$-dimensional sphere (or hyperboloid). On the
other hand, Gaudin's magnet on 3 sites generates continuous quasi-exactly
solvable hamiltonians \cite{Ush}. Considered as quadratic
forms in generators of $sl(2)$ such a hamiltonian determines a separated
coordinate system on the 2-sphere (or hyperboloid). The non-equivalent
separated systems correspond \cite{WintYaF} to $SL(2)$-orbits (under the
adjoint action). The quadratic forms in generators of $U_q(sl(2))$
might have a similar relation to hypothetical "separated coordinate
systems" on quantum spheres and hyperboloids.
\medskip

\section*{Acknowledgements}

Some results of this paper were presented in author's talk at the Vth
International Conference on Mathematical Physics, String Theory and
$2d$ Gravity (Alushta, June 1994).
It is a pleasure to thank the organizers and participants for the nice
atmosphere. I am grateful to P.Wiegmann for collaboration and numerous
discussions. This work was partially supported by grant 93-02-14365 of the
Russian Foundation of Fundamental Research, by ISF grant MGK000 and by
ISTC grant 015.

\end{document}